\documentclass[a4paper]{jpconf}
\usepackage{iopams}
\usepackage{cite}

\usepackage[dvips]{graphicx}
\usepackage{pstricks}

\usepackage{epsfig}

\global\arraycolsep=2pt

\begin{document}
\title{Lateral Casimir forces on parallel plates and concentric cylinders with corrugations.}
\author{In\'{e}s Cavero-Pel\'{a}ez}
\address{Laboratoire Kastler Brossel,
Universit\'{e} Pierre et Marie Curie, ENS, CNRS, Campus Jussieu, 
University Paris 6, Case 74, F-75252 Paris, cedex 05, France.}
\ead{cavero@spectro.jussieu.fr}
\author{Kimball A. Milton
, Prachi Parashar and K. V. Shajesh
}
\address{Oklahoma Center for High Energy Physics
and Homer L. Dodge Department of Physics and Astronomy,
University of Oklahoma, Norman, OK 73019, USA.}
\ead{milton@nhn.ou.edu, prachi@nhn.ou.edu, shajesh@nhn.ou.edu}

\begin{abstract}
In this paper we are giving a quantitative description of two different configurations for noncontact gears. We consider the solutions from a perturbative calculation for two semitransparent parallel plates and concentric cylinders both with corrugations on the inner surfaces. In the case of corrugated parallel plates we discuss results from first- and second-order perturbation calculation in the corrugation amplitudes and we will concentrate on the first-order perturbation for the case of the corrugated concentric cylinders (the second order calculation is under study), both for the weak and strong couplings. We compare the perturbative results with the results from the PFA and an exact weak coupling calculation. 

\end{abstract}


\section{Introduction.}

When two parallel plates with corrugations are situated close to each other, besides the usual Casimir attraction that takes place between them \cite{casimir48}, there exists another contribution to the attraction that depends on the corrugations of the surface of the plates and therefore, on the  non-normal coordinates (in the case of the parallel plates this coordinate is along the parallel surfaces where the corrugations are imposed, and in the case of the concentric cylinders this coordinate is the azimuthal angle with origin at the center of the cylinders). This new term gives rise to a lateral force or torque that could, in principle, make one plate slide with respect to the other or one cylinder rotate with respect to the other. Here we discuss such a situation. We base our discussion on a perturbative calculation details of which can be found in \cite{gearsI,gearsII,AFcas}. The case of QED parallel plates has been worked out in the perturbative regime in \cite{emig2003, rodrigues2006}.  Some nanoscale mechanical devices based on the lateral Casimir force between different geometries have been proposed in \cite{ashourvan-miri-gloestanian, lombardo-mazzi-villar, miri-golestanian}.

We divide the paper in two relevant parts. The first one is devoted to corrugated parallel plates. We show and discuss the perturbative results for two limiting cases, strong and weak-coupling. The weak-coupling limit  is also compared with an exact calculation and the PFA solution. The second part of the paper shows the results for corrugated concentric cylinders. We give perturbative results for both the Dirichlet and the weak-coupling limits. We extrapolate these results to the case of very large radii of the cylinders while keeping constant the distance between them. This situation corresponds to parallel plates.


\section{Parallel plates with corrugations.} 

Let's consider two semitransparent plates lying on the $x$-$y$ plane and placed parallel to each other. The plates have corrugations on the facing surfaces along the $y$-axes. We consider translational invariance on the $x$-axes and we work at a given frequency. The average distance between the parallel slabs is $a$ and we  model the corrugations by the functions $h_1(y)$ and $h_2(y)$. These are defined such that at any point the inequality $a\ge h_1(y)-h_2(y)$ holds.  

\begin{figure}
\begin{center}
\includegraphics[width=70mm]{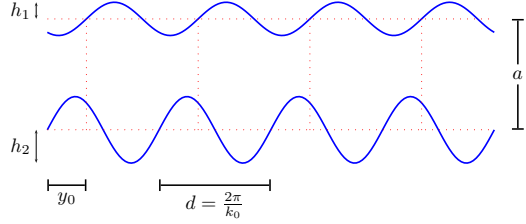}
\caption{Parallel plates with sinusoidal corrugations.}
\label{ppcorrugationss}
\end{center}
\end{figure}

For the cases studied here we choose sinusoidal corrugations, $h_i(y)=h_i\sin (k_i y)$. Each plate is described by a potential of the form,
\begin{equation}
V_i(z,y) = \lambda_i \,\delta (z - a_i - h_i(y)),\label{potential}
\end{equation}
where $i=1,2$ are labels that identify the individual plates. In this way we can consider our system as two semitransparent parallel plates making up the background and described by the potentials $V_i^{(0)}(z)=\lambda_i\delta(z-a_i)$ for $i=1, 2$. On the top of it we have the corrugations $h_i(y)$ which can be considered to be a perturbation of the background (see \fref{ppcorrugationss}).

Let's make one of the plates shift a distance $y_0$ with respect to the other one.
\numparts
\begin{eqnarray}
h_1(y) &=& h_1 \sin [k_{0}(y + y_0)],
\\
h_2(y) &=& h_2 \sin [k_{0} y],
\end{eqnarray}
\endnumparts
where $k_0 = 2\pi/d$ is the wavenumber corresponding to the corrugation wavelength $d$ \footnote{The same wavelength is assumed in both plates because it is only in this case that we find a leading order contribution to the lateral force.}. This gives rise to a force in the lateral direction, the so-called lateral Casimir force,
\begin{equation}
F_{\rm {Lat}} = - \frac{\partial E}{\partial y_0}.
\label{lateralCF}
\end{equation}

Since the corrugations can be considered a perturbation over the background $V_i^{(0)}(z)$, we can easily extract their contribution by calculating the deviation of the total potential $V_i(z,y)$ from that of the background,
\begin{eqnarray}
\Delta V_i(z,y)&=&V_i(z,y)-V_i^{(0)}=\lambda_i \,\delta (z - a_i - h_i(y))-\lambda_i\delta(z-a_i)\nonumber\\
&=& \sum_{n=1}^\infty \frac{[-h_i(y)]^n}{n!}
\frac{\partial^n}{\partial z^n} V_i^{(0)}(z)
=  \sum_{n=1}^\infty V_i^{(n)}(z,y).
\end{eqnarray}

The key formula to calculate the Casimir energy is \cite{milton-mscattering2008,kenneth2006,emig2008},
\begin{equation}
\Delta E = \frac{i}{2 \tau} \,\rm{Tr} \,\rm{ln} \,G {G^{(0)}}^{-1},
\label{deltaE}
\end{equation}
where the Green's function $G$ satisfies a differential equation with a potential function that describes the corrugations in both semitransparent plates,
\begin{equation}
\Big[ -\partial^2 + V_1 + V_2 \Big] G = 1,
\end{equation}
and the corresponding Green's function associated with the 
background obeys the differential equation
\begin{equation}
\Big[ -\partial^2 + V_1^{(0)} + V_2^{(0)} \Big] G^{(0)} = 1,
\label{G0eqn}
\end{equation}
where the corrugations are not present.
However, since only the term containing the perturbative potential in both plates play a role in determining the lateral Casimir force, we can reduce the above more general formula \eref{deltaE} to 
\begin{equation}
E_{12} = - \frac{i}{2\tau} \,\rm{Tr} \,\rm{ln}
\Big[ 1 - G_1 \Delta V_1 G_2 \Delta V_2 \Big],
\label{lateralCE}
\end{equation}
where now $G_i$ refers to the Green's function when only one of the plates has corrugations on it. It obeys the equation
\begin{equation}
\left[-\partial^2 + V_1^{(0)} + V_2^{(0)} + \Delta V_i\right]G_i = 1.
\end{equation}
It is worth noticing that only the corrugations from plate $i$ enter this equation.

Expanding the above formula \eref{lateralCE} in a second- and fourth-order perturbative calculation in the corrugation amplitudes, gives expressions for the leading- and next-to-leading-order lateral Casimir energy. The general formulas turn out to be very long and complicated and they will not be shown here (details of the calculation as well as final formulas can be found in \cite{gearsI, AFcas}); here we focus on the solutions for the Dirichlet and weak coupling limits.


\subsection{Dirichlet limit.}
Here we consider the strong coupling limit, where $a\lambda_{1,2}\gg 1$, both in the leading and next-to-leading orders.

\subsubsection{Leading order.}
\begin{figure}
\begin{center}
\includegraphics[width=80mm]{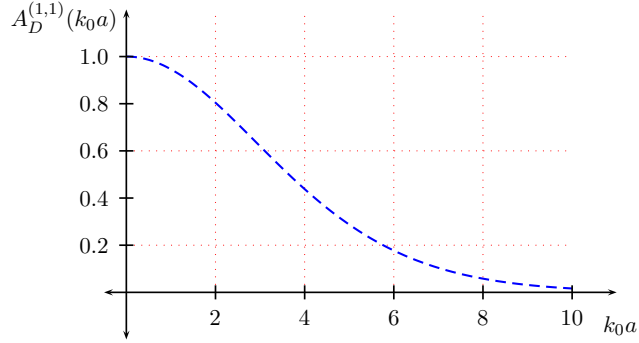}
\caption{Dirichlet limit: Plot of $A^{(1,1)}_D(k_0a)$ versus $k_0a$. }
\label{A11-D-versus-t0.ps}
\end{center}
\end{figure}

Expanding \eref{lateralCE} to second order gives us the first non-zero contribution to the interaction energy,
\begin{equation}
\frac{E_{12}^{(2)}}{L_xL_y}
= \cos (k_0 y_0) \,\frac{\pi^2}{240\,a^3} \,\frac{h_1}{a}\frac{h_2}{a}
\, A^{(1,1)}_D(k_0a),
\label{E12-2-p}
\end{equation}
where the function $A^{(1,1)}_D(k_0a)$ has the form
\begin{equation}
A^{(1,1)}_D(t_0) = \frac{15}{\pi^4} \int_0^\infty \bar{s} \,d\bar{s}
\int_{-\infty}^{\infty} dt \frac{s}{\sinh s} \frac{s_+}{\sinh s_+},
\end{equation}
and it is normalized such that 
$A^{(1,1)}_D(0) = 1$, which becomes a convenience when we compare the 
results with those obtained by the proximity force approximation. We have used the notation $s^2 = \bar{s}^2 + t^2$,
$s^2_\pm = \bar{s}^2 + (t \pm t_0)^2$, and $t_0=k_0a$.
This expression can be numerically evaluated and has been plotted in \fref{A11-D-versus-t0.ps}.

The lateral Casimir force can be calculated using 
\eref{lateralCF} which yields
\begin{equation}
F_{\rm{Lat,D}}^{(2)} 
= 2 \,k_0a \,\sin (k_0y_0) \,\left| F_{\rm{Cas,D}}^{(0)} \right| 
\,\frac{h_1}{a} \frac{h_2}{a} \, A^{(1,1)}_D(k_0a),
\label{latCF2}
\end{equation}
where $|F_{\rm{Cas,D}}^{(0)}|$ 
is the magnitude of the normal scalar Casimir force
between two uncorrugated parallel Dirichlet plates given as
\begin{equation}
\frac{F_{\rm{Cas,D}}^{(0)}}{L_xL_y}
= - \frac{\pi^2}{480} \frac{1}{a^4}.
\end{equation}

\subsubsection{Next-to-leading order.}
The third-order term in the perturbation expansion gives zero contribution in the particular case when the wavelength is the same in both plates. If we do not restrict ourselves to this condition, the third order would give a finite contribution to the interacting energy becoming, in this case, the leading contribution itself\footnote{because in such a case the second order expansion exactly cancels.}. Since we are assuming $k_1=k_2=k_0$, we expand \eref{lateralCE} up to fourth order to get the next-to-leading-order contribution to $E_{12}$. In the strong-coupling limit this gives us
\begin{eqnarray}
\frac{E_{12}^{(4)}}{L_xL_y}
&=& \frac{\pi^2}{240\,a^3} \frac{h_1}{a} \frac{h_2}{a} \frac{15}{4}
\Bigg[ \,\cos (k_0y_0) 
\Bigg\{ \frac{h_1^2}{a^2} \,A^{(3,1)}_D(k_0a)
+ \frac{h_2^2}{a^2} \,A^{(1,3)}_D(k_0a)
\Bigg\}\nonumber\\
&&- \cos (2k_0y_0) \,\frac{1}{2} \frac{h_1}{a} \frac{h_2}{a} \,A^{(2,2)}_D(k_0a)
\Bigg], 
\end{eqnarray}
where we have introduced the functions,
\numparts
\begin{eqnarray}
A^{(3,1)}_D(t_0) &=&A^{(1,3)}_D(t_0)\nonumber\\
&=& \frac{1}{2\pi^4} \int_0^\infty \bar{s}\,d\bar{s} 
\int_{-\infty}^{\infty} dt 
\frac{s}{\sinh s} \frac{s_+}{\sinh s_+}
\left[ 4 \frac{s}{\tanh s} \frac{s_-}{\tanh s_-}
+ 2 \frac{s}{\tanh s} \frac{s_+}{\tanh s_+} - s^2 - s_-^2 \right],\nonumber\\
\\
A^{(2,2)}_D(t_0) &=& \frac{1}{\pi^4} \int_0^\infty \bar{s}\,d\bar{s} 
\int_{-\infty}^{\infty} dt
\left[ \frac{s^2}{\sinh^2 s} \frac{s_-^2}{\sinh^2 s_-}
+ 2 \frac{s^2}{\tanh^2 s} \frac{s_+}{\sinh s_+} \frac{s_-}{\sinh s_-}
\right].
\end{eqnarray}\endnumparts
\begin{figure}
\begin{tabular}{cc}
\includegraphics[width=80mm]{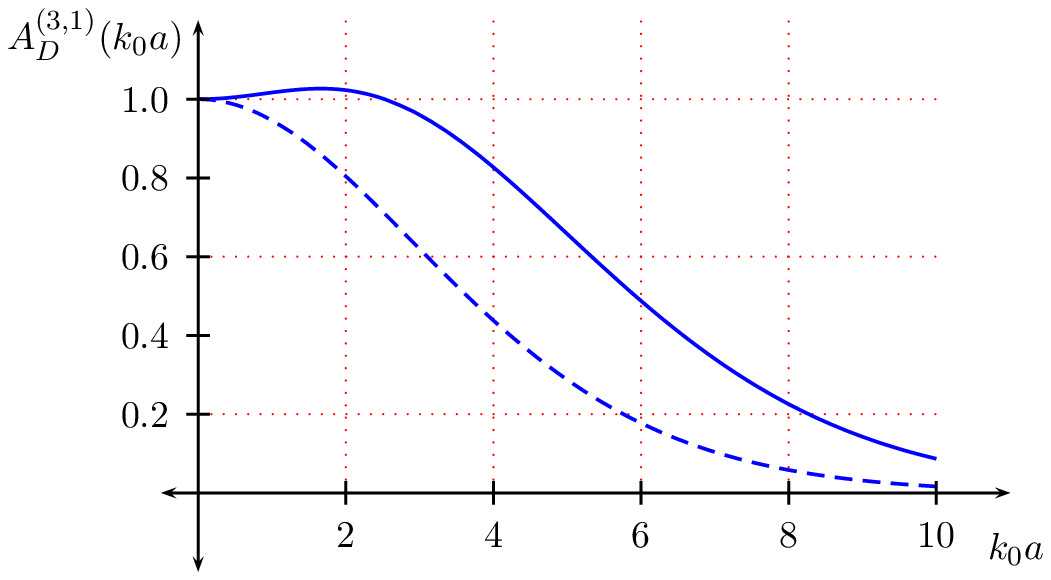} &
\includegraphics[width=80mm]{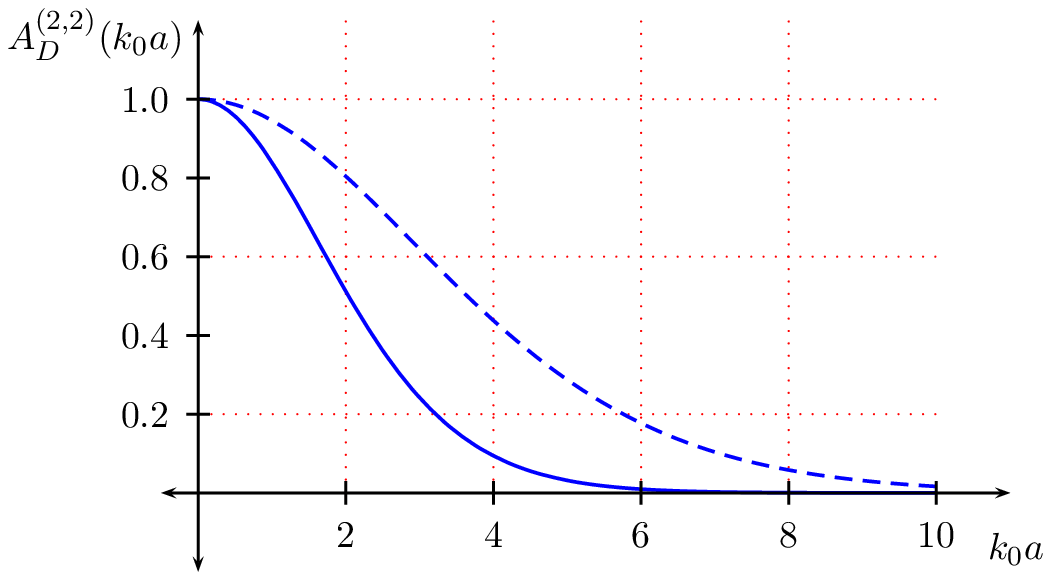}
\end{tabular}
\caption{
Dirichlet limit:
Plot of $A^{(3,1)}_D(k_0a)$ and $A^{(2,2)}_D(k_0a)$
versus $k_0a$. The dashed curve represents $A^{(1,1)}_D(k_0a)$
which is plotted here for reference. }
\label{AD3122-fig}
\end{figure}
These functions can be calculated numerically and they are plotted in \fref{AD3122-fig} for a later discussion.

The next-to-leading-order contribution to the lateral Casimir force 
in the Dirichlet limit reads
\begin{eqnarray}
F_{\rm{Lat,D}}^{(4)}
&=& 2 \,k_0a \, \sin (k_0y_0) \,\left|F_{\rm{Cas,D}}^{(0)} \right| 
\,\frac{h_1}{a} \frac{h_2}{a} \,\frac{15}{4}
\Bigg[ \left( \frac{h_1^2}{a^2} + \frac{h_2^2}{a^2} \right) A^{(3,1)}_D(k_0a)\nonumber\\
&&- 2 \cos (k_0y_0) \,\frac{h_1}{a} \frac{h_2}{a} \,A^{(2,2)}_D(k_0a) \Bigg].
\label{latCF4}
\end{eqnarray}

\subsubsection{Discussion.}
In order to compare our results, we are going to confine ourselves to the case where the amplitudes of the corrugations in both plates are the same, $h_1=h_2=h$. Also, by observing equations (\ref{latCF2}) and (\ref{latCF4}) we notice that there is a common factor in all of them. It involves the normal Casimir force, as well as terms dependent on the geometry of the corrugations and on the initial shift $y_0$ between the plates. By removing this common factor, what is left is just a dimensionless weight factor. We can therefore normalize the above expressions to the common factor and compare the resulting dimensionless term that we obtain in this manner from each of the mentioned formulas. In this spirit we define the weight factors as
\begin{equation}
{\cal F}_{D} \left( k_0a,\frac{h}{a},k_0y_0 \right)
= \frac{F_{\rm{Lat,D}} \left(k_0a,\frac{h}{a},k_0y_0 \right)}
       {2 \, \left| F^{(0)}_{\rm{Cas,D}} \right| 
        \,\frac{h^2}{a^2} \,k_0a \,\sin (k_0y_0)}.
\end{equation}
We shall use superscripts $[m,n]$ on these dimensionless
quantities to denote the order in the power series expansion
to which the lateral force (not $\cal{F}$) has been calculated. 
In particular $m$ will signify the order in the 
parameter $k_0a$, and $n$ will denote the order in $h/a$.
The perturbative results are complete
in the $k_0a$ dependence and thus will be denoted by 
${\cal F}^{[\infty, n]}$. 
In these notations our results can be summarized as
(see equations \eref{latCF2} and \eref{latCF4})
\begin{equation}
{\cal F}_{D}^{[\infty,4]}
= A^{(1,1)}_D(k_0a) +\frac{15}{2} \frac{h^2}{a^2} 
\Big[ A^{(3,1)}_D(k_0a) - \cos (k_0y_0) \,A^{(2,2)}_D(k_0a) \Big],
\label{weight-factor-D}
\end{equation}
where ${\cal F}_{D}^{[\infty , 4]}$ means weight factor of the lateral Casimir force up to the fourth order. That is, it includes both leading and next-to-leading orders. 

\begin{figure}
\begin{tabular}{cc}
\includegraphics[width=80mm]{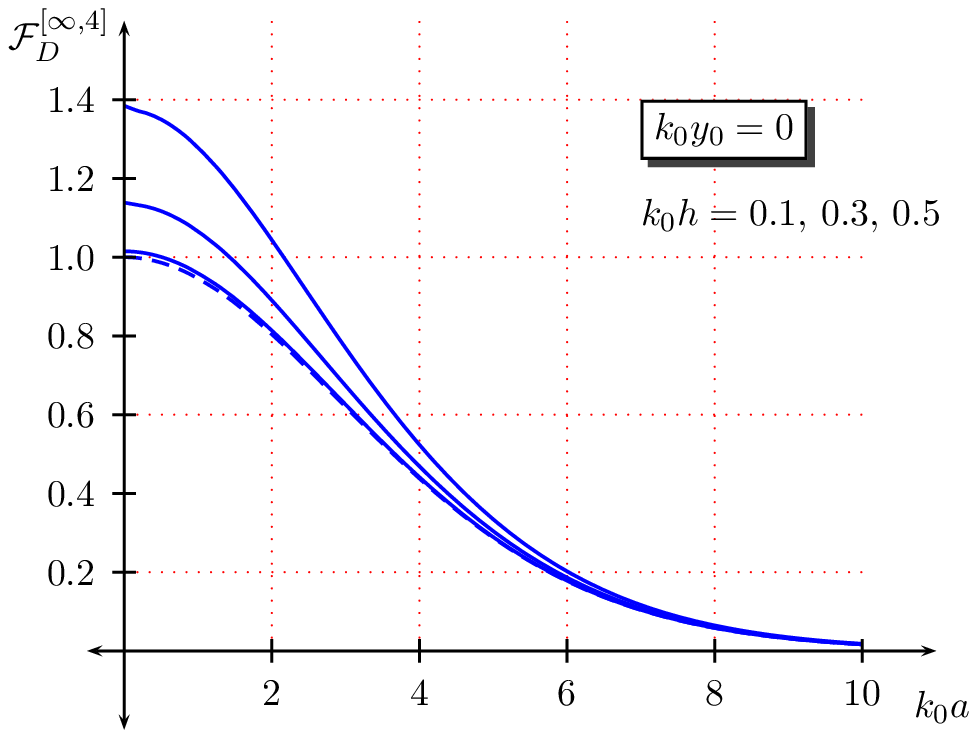} &
\includegraphics[width=80mm]{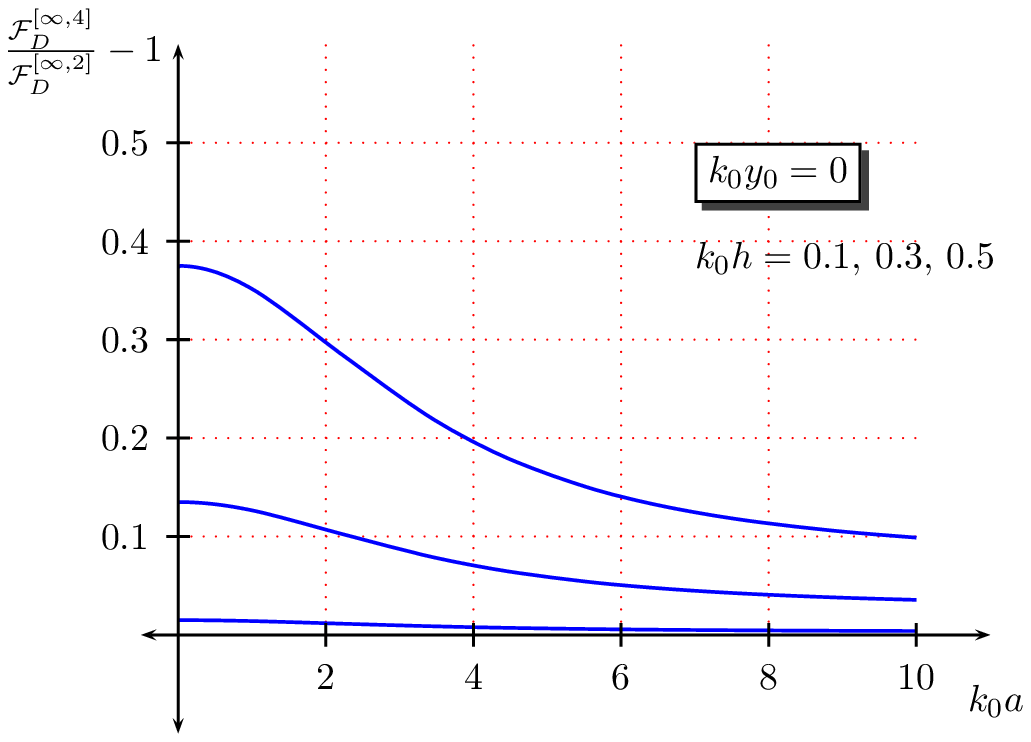}
\\
\includegraphics[width=80mm]{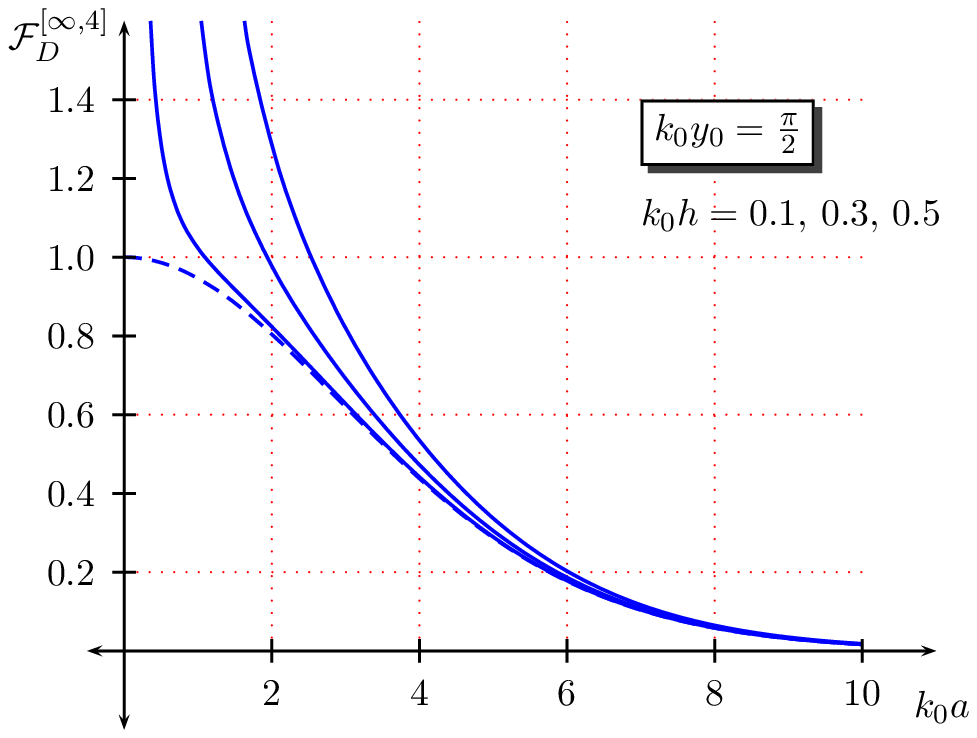} &
\includegraphics[width=80mm]{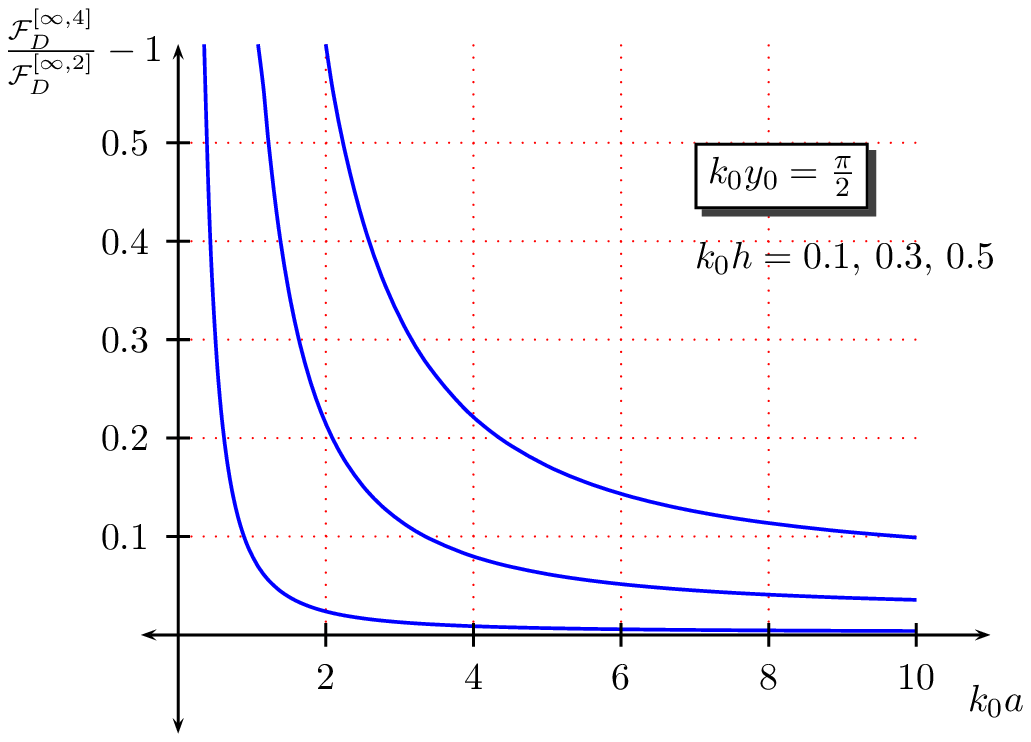}
\\
\includegraphics[width=80mm]{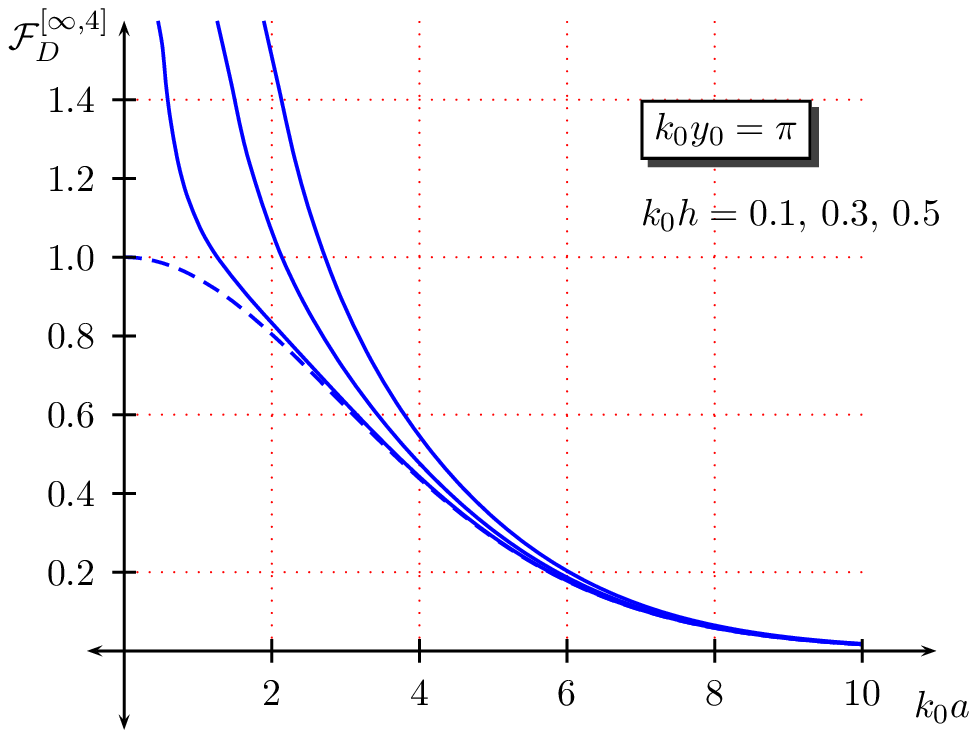} &
\includegraphics[width=80mm]{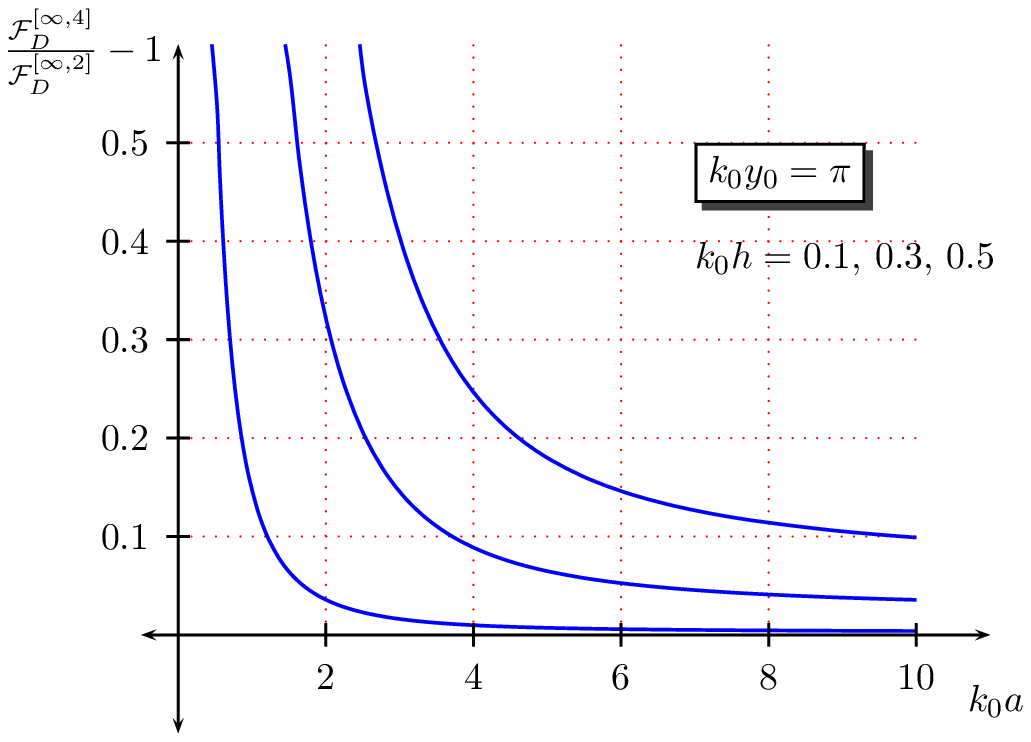}
\end{tabular}
\caption{ 
Dirichlet limit:
The plots on the left show ${\cal F}_D^{[\infty,4]}$ versus $k_0a$ for various
values of $k_0y_0$. The dashed curve represents ${\cal F}_D^{[\infty,2]}$
which is plotted here for reference.
In each plot the higher values of $k_0h$ deviate more from the reference.
The plots on the right show the fractional correction in the
next-to-leading-order contribution.
In each plot the higher values of $k_0h$ have larger corrections.  }
\label{leading2ntol}
\end{figure}

We can in fact look at how much the next-to-leading order contributes to the force compared to the leading order. The left plots in \fref{leading2ntol} show the second-order contribution to the force with a dashed line and the fourth-order contribution with a solid line for different values of the angle and constant values of $k_0h$. We see that for a fixed offset $k_0y_0$ the fourth order contribution becomes more important for high $k_0h$ that means, for bigger corrugations. We show the plot for $k_0y_0=0$ to illustrate this fact more clearly because in this case the weight factor does not diverge for small $k_0a$. However, at that value of the offset, the lateral force is zero as it can be noticed straightforwardly from \eref{latCF2} and \eref{latCF4}. For any other value of the offset, there are points where the graph diverges. This is because as $k_0h$ approaches $2k_0a$, the expansion is not longer valid. It corresponds to the case when the distance between the plates is smaller than the sum of the corrugations on both plates.

On the right hand side of the same figure we plot the fractional correction of the next-to-leading-order contribution relative to the leading-order. It shows that at any given offset $k_0y_0$ and $k_0h$ the fractional error is higher for small separations and big corrugations give higher next-to-leading-order relative errors.

In the absence of an exact answer, for the Dirichlet case, 
for comparison is not possible 
to extract the precise error in the perturbative results. 
In the next section we shall evaluate the corresponding result in 
the weak coupling limit and evaluate the lateral Casimir force 
non-perturbatively. We show that the error in the lateral force,
for the weak coupling limit,
after the next-to-leading order has been included is 
small for $k_0h \ll 1$. Presuming that the same would
hold for the Dirichlet case, we can expect the error 
to be sufficiently small to use the fourth-order results for comparison with experimental data. 


\subsection{Weak coupling limit.}
We consider now $a\lambda_{1,2}$ to be very small and we study the second and fourth order in the perturbation series.

\subsubsection{Leading order.}
\begin{figure}
\begin{center}
\includegraphics[width=80mm]{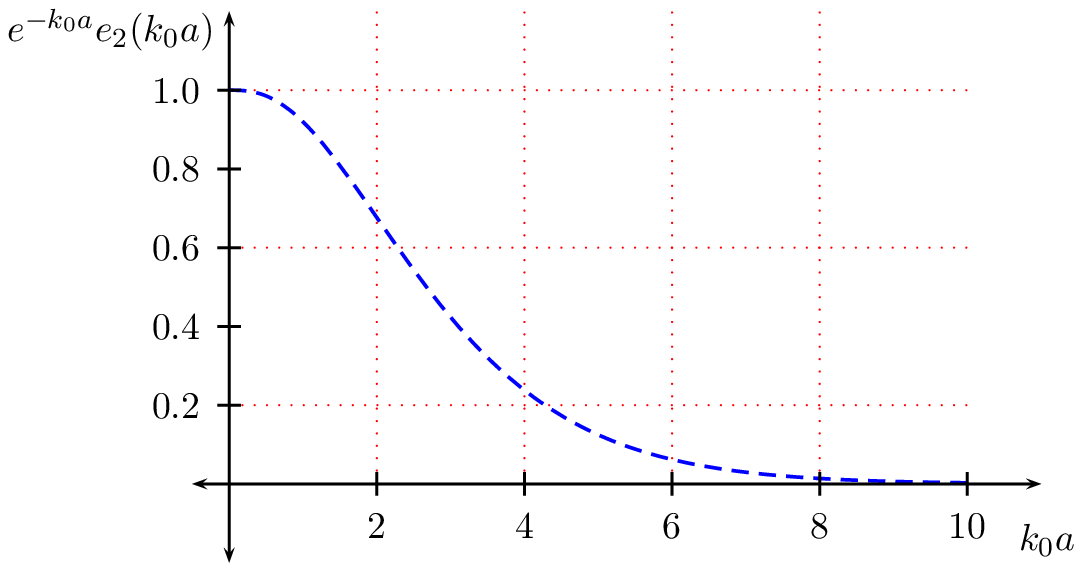}
\caption{Weak coupling limit:
Plot of $A^{(2)}_W(k_0a) = e^{-k_0a} e_2(k_0a)$ versus $k_0a$. }
\label{A2-W-t0-fig}
\end{center}
\end{figure}
In this case the expansion of \eref{lateralCE} leads to an expression that can be evaluated analytically
\begin{equation}
E_{\rm{12,W}}^{(2)}
= \cos (k_0y_0) \,\left| E_{\rm{Cas,W}}^{(0)} \right| 
\,\frac{h_1}{a} \frac{h_2}{a} \, A^{(2)}_W(k_0a),
\end{equation}
where we have introduced the function
\begin{equation}
A^{(2)}_W(t_0) = \frac{t_0^3}{2} \,\frac{\partial^2}{\partial t_0^2}
\bigg[ \frac{1}{t_0} \,e^{-t_0} \bigg]
= e^{-t_0} \sum_{m=0}^2 \, \frac{t_0^m}{m!} = \frac{e_2(t_0)}{e^{t_0}},
\end{equation}
plotted in \fref{A2-W-t0-fig}.
We evaluate the lateral Casimir force to this perturbation order as 
\begin{equation}
F_{\rm{Lat,W}}^{(2)}
= k_0a \,\sin (k_0y_0) \,\left| F_{\rm{Cas,W}}^{(0)} \right| 
\,\frac{h_1}{a} \frac{h_2}{a} \, A^{(2)}_W(k_0a).
\label{latf-2W}
\end{equation}

\subsubsection{Next-to-leading order.}
We can write the total fourth-order contribution to the interaction
energy in the presence of corrugations in the weak limit to be
\begin{equation}
E_{12,W}^{(4)}
= \left| E^{(0)}_{\rm{Cas,W}} \right| \frac{h_1}{a} \frac{h_2}{a} \frac{3}{2}
\left[ \cos (k_0y_0) 
\left( \frac{h_1^2}{a^2} + \frac{h_2^2}{a^2} \right) A^{(4)}_W(k_0a)
- \cos (2k_0y_0) \frac{1}{2} \frac{h_1}{a} \frac{h_2}{a} A^{(4)}_W(2k_0a)
\right],
\label{DE12-4W}
\end{equation}
where in contrast to the result in the Dirichlet limit we observe
the appearance of the same function, though with different arguments,
as coefficients. This function has been suitably defined as
\begin{equation}
A^{(4)}_W(t_0) = \frac{t_0^5}{4!} \frac{\partial^4}{\partial t_0^4}
\bigg[ \frac{1}{t_0} \, e^{-t_0} \bigg]
=e^{-t_0} \sum_{m=0}^4 \, \frac{t_0^m}{m!} = \frac{e_4(t_0)}{e^{t_0}},
\end{equation}
\begin{figure}
\begin{tabular}{cc}
\includegraphics[width=80mm]{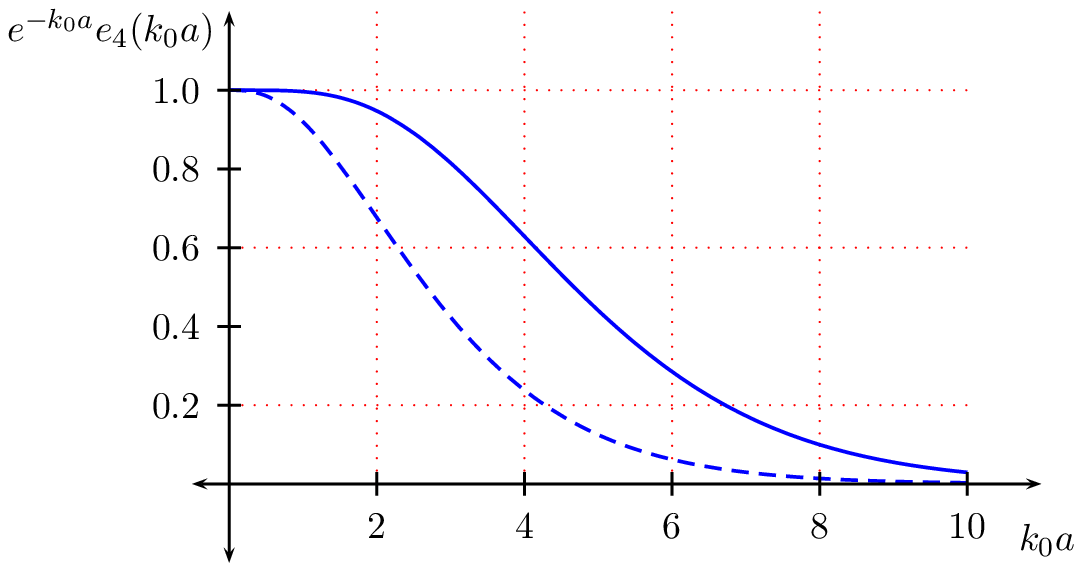} &
\includegraphics[width=80mm]{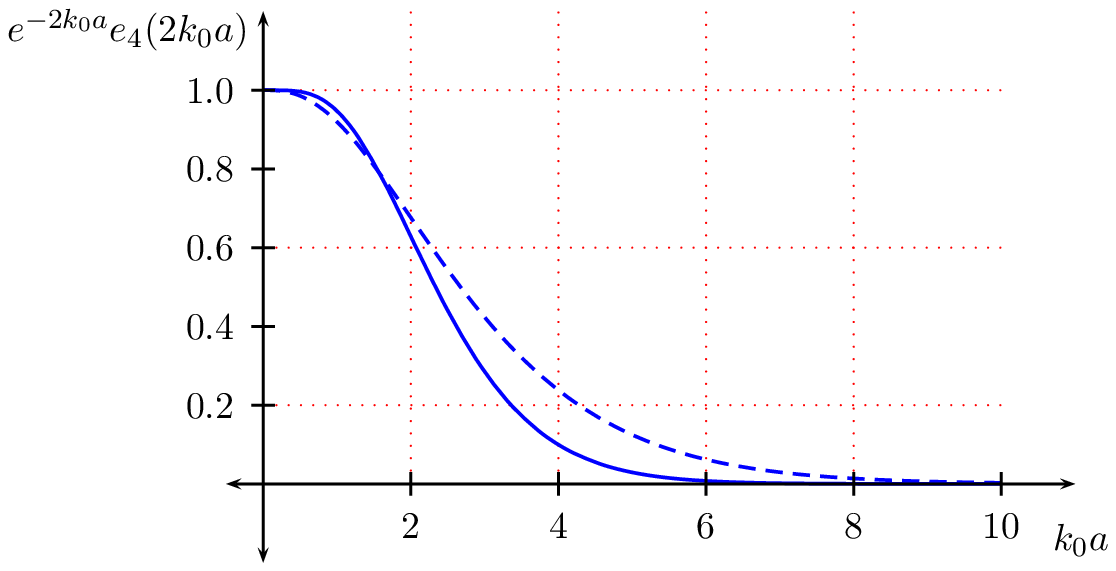}
\end{tabular}
\caption{
Weak coupling limit:
Plot of $A^{(4)}_W(k_0a)$ and $A^{(4)}_W(2k_0a)$
versus $k_0a$. The dashed curve represents $A^{(2)}_W(k_0a)$
which is plotted here for reference. }
\label{A4-W-t0-fig}
\end{figure}
so that it equals unity at $t_0=0$. 
These are plotted in \fref{A4-W-t0-fig}.
Using \eref{DE12-4W} in \eref{lateralCF} we find that the fourth-order
contribution to the lateral Casimir force in the weak coupling limit
equals
\begin{equation}
F_{\rm{Lat,W}}^{(4)}
= k_0a \, \sin (k_0y_0) \,\left|F_{\rm{Cas,W}}^{(0)} \right|
\,\frac{h_1}{a} \frac{h_2}{a} \,\frac{3}{2}
\left[ \left( \frac{h_1^2}{a^2} + \frac{h_2^2}{a^2} \right) A^{(4)}_W(k_0a)
- 2 \cos (k_0y_0) \,\frac{h_1}{a} \frac{h_2}{a} \,A^{(4)}_W(2k_0a) \right].
\label{latf-4W}
\end{equation}

\subsubsection{Proximity force approximation.}
It is interesting to find this result since, as we shall see later, the weak-coupling limit can be solved exactly. That will allow us to compare both the perturbative and the proximity force approximation results to the exact solution.
For sufficiently small distances $a$ (in comparison to $d=2\pi \,k_0^{-1}$) 
we can treat the plates to be built out of small sections for which the energy is approximately that of the parallel plate 
geometry. This is the so-called proximity force approximation. To compute the force under this approximation we consider the distance between the plates to be
\begin{equation}
a(y) = a + h_2 \sin[k_0 y] - h_1 \sin [k_0(y + y_0)],
\end{equation}
so that we can integrate the resulting energy after substituting the above in the expression for the Casimir energy between two plates in the weak coupling limit.

Therefore, in the PFA, the lateral Casimir force in the weak coupling limit reads,
\begin{equation}
F^{\rm{PFA}}_{\rm{Lat,W}}
= k_0a \,\sin (k_0y_0) \,\left| F^{(0)}_{\rm{Cas,W}} \right|
\,\frac{h_1}{a} \frac{h_2}{a} \, \frac{1}{(1 - \frac{r^2}{a^2})^\frac{3}{2}},
\end{equation}
for $|h_1| + |h_2| < a$.

\subsubsection{Nonperturbative result.}
\begin{figure}
\begin{tabular}{cc}
\includegraphics[width=80mm]{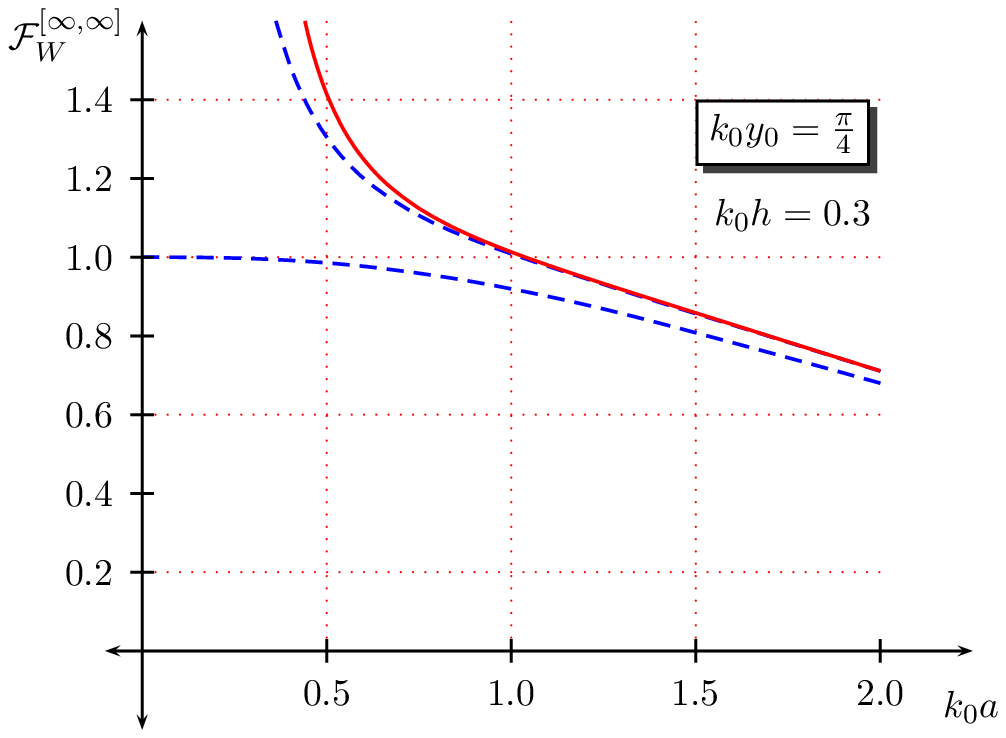} &
\includegraphics[width=80mm]{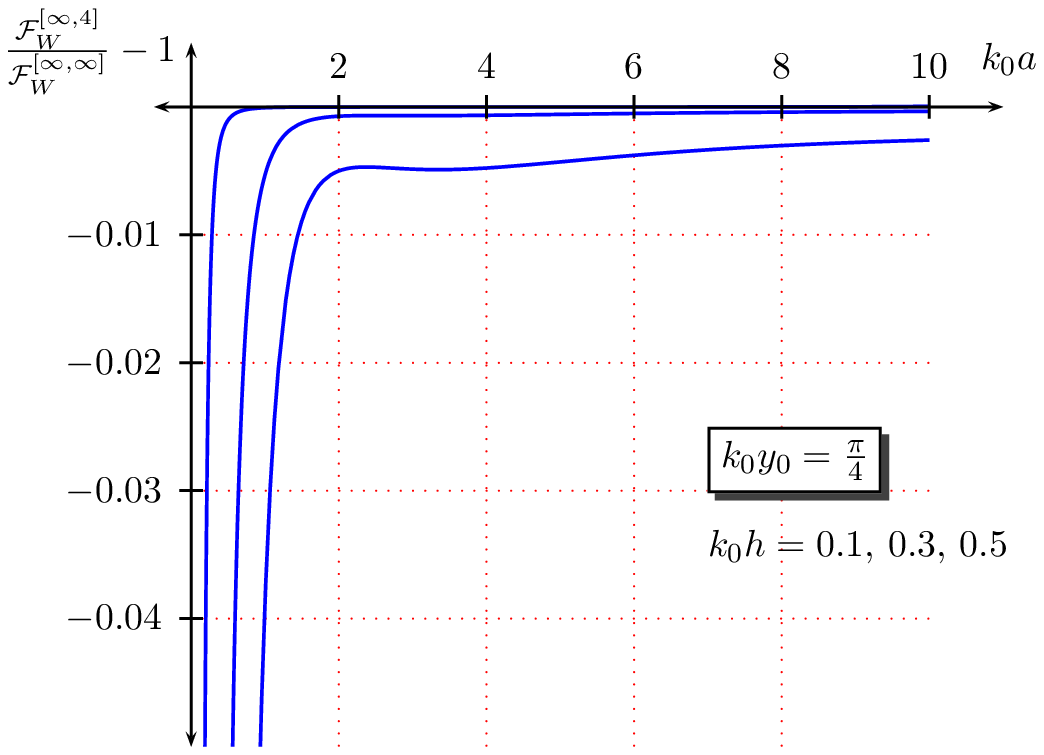}
\\
\includegraphics[width=80mm]{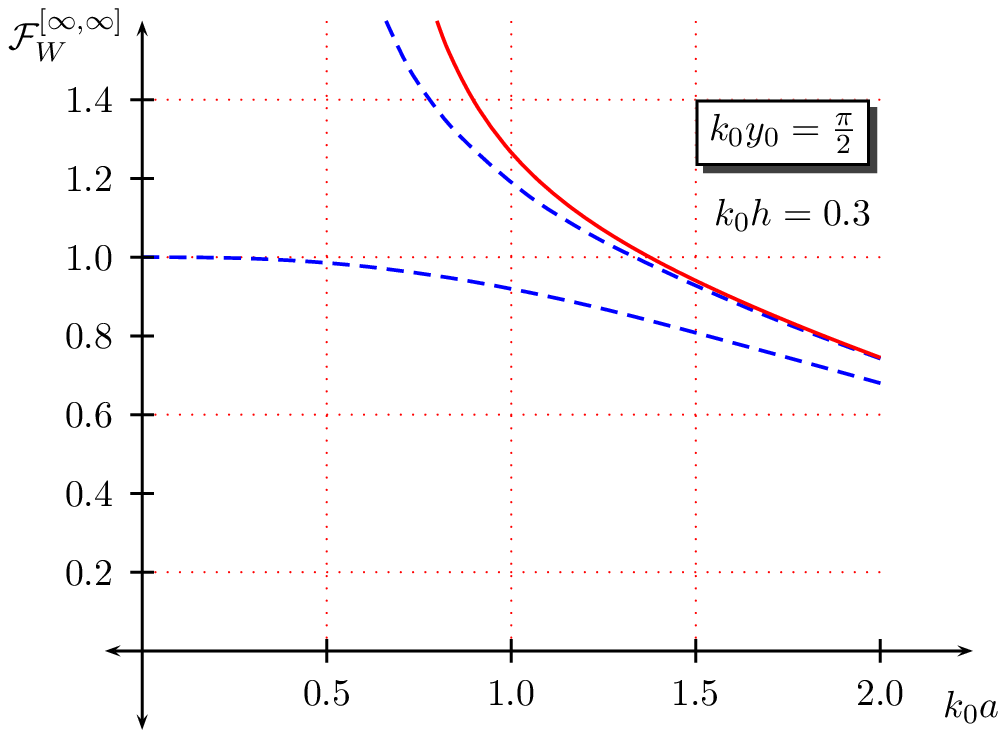} &
\includegraphics[width=80mm]{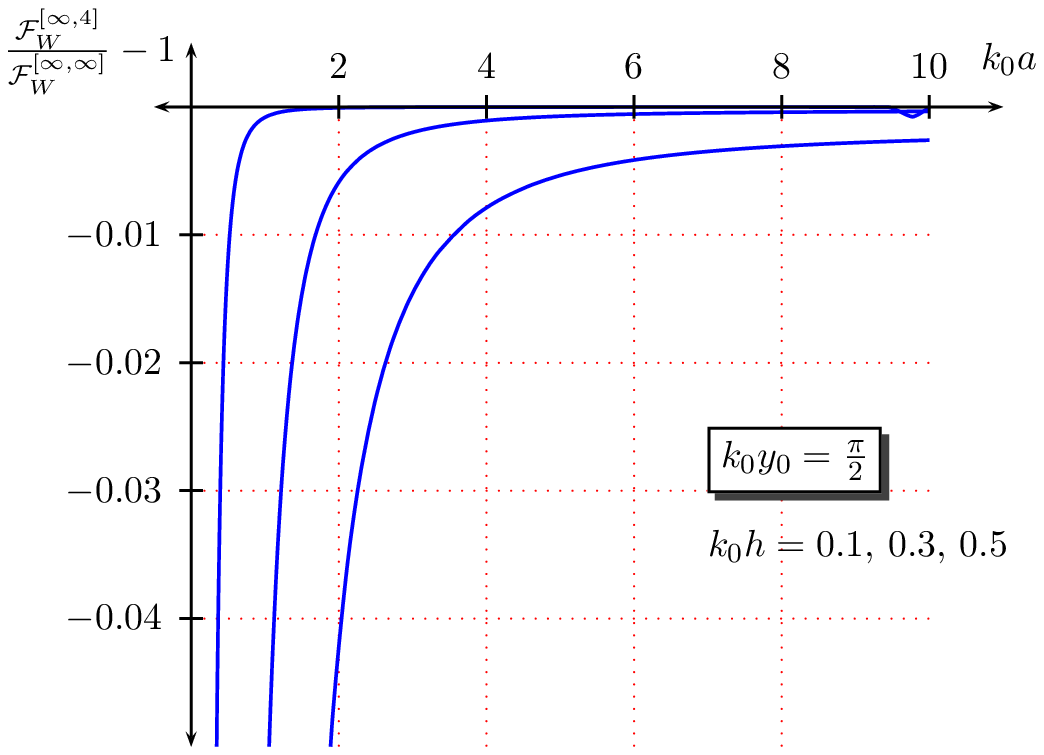}
\\
\includegraphics[width=80mm]{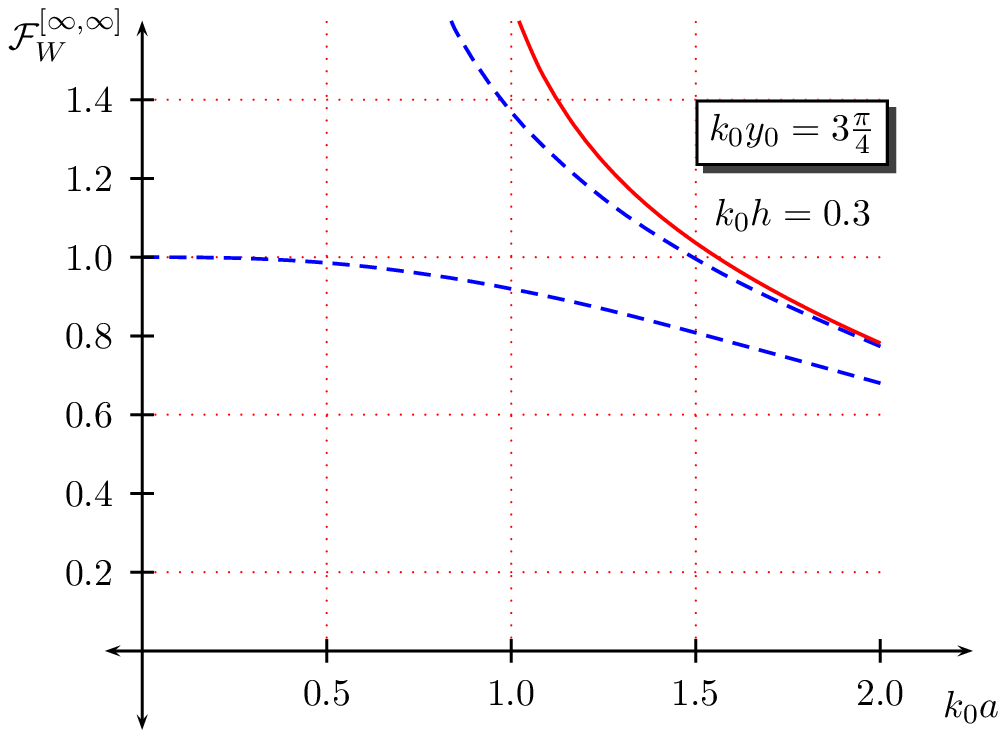} &
\includegraphics[width=80mm]{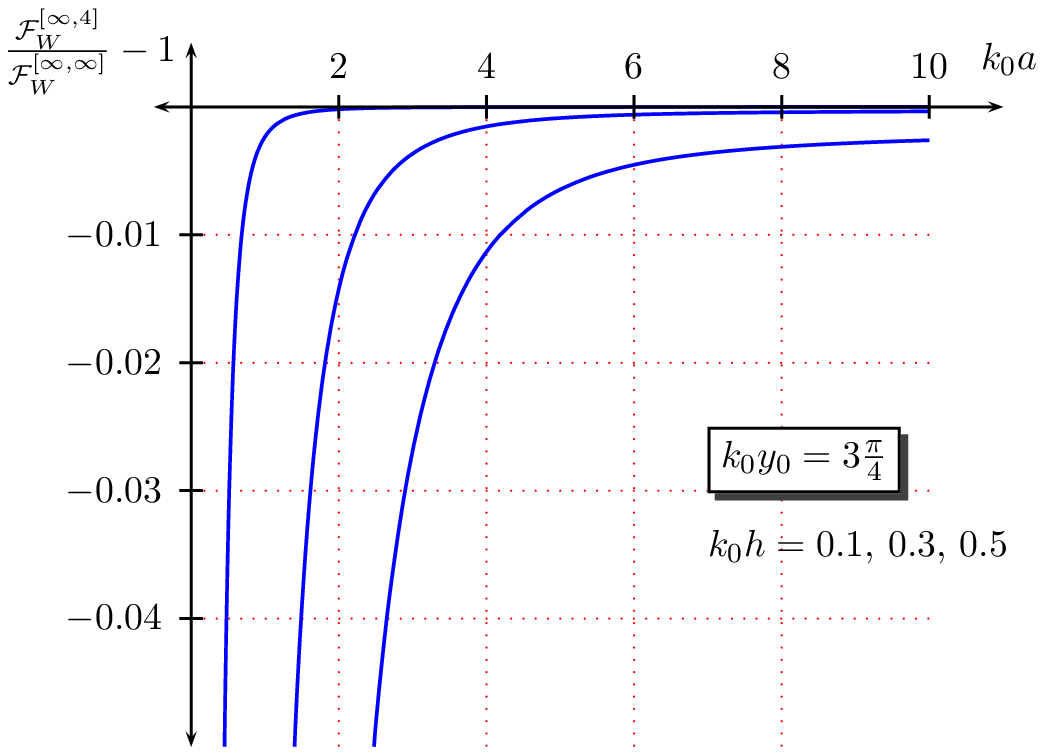}
\end{tabular}
\caption{
Weak coupling limit:
The plots on the left show ${\cal F}_D^{[\infty,\infty]}$ versus $k_0a$ 
for various values of $k_0y_0$ at $k_0h=0.3$.
The dashed curves represents
${\cal F}_D^{[\infty,2]}$ and ${\cal F}_D^{[\infty,4]}$
which is plotted here for reference.
In each plot the exact value is greater than the value estimated
by ${\cal F}_D^{[\infty,4]}$.
The plots on the right show the fractional error in the perturbative result.
In each plot the higher values of $k_0h$ have larger errors.  }
\label{exact-W-fig}
\end{figure}
It was shown in~\cite{milton-mscattering2008} that exact results for the Casimir force,
in the weak coupling limit, can be achieved for specific geometries.
This is being extended for a class of geometries,
for the scalar case, in~\cite{milton-wagner2008}, and for dielectrics in electromagnetism~\cite{milton-parashar-wagner08}. 
In our case the solution of the exact weak coupling reads
\begin{equation}
\frac{F_{\rm{Lat,W}}}{L_xL_y} = - \frac{\lambda_1 \lambda_2}{32\pi^2a^2}
\frac{h_1}{a} \frac{h_2}{a} (k_0a)^4
\frac{1}{\pi} \int_{-\infty}^\infty \frac{d t}{t}
\,{\rm Re} \bigg[ \frac{\sin (t + k_0y_0)}
            {[(t + \rmi k_0a)^2 + \{k_0 r(t)\}^2]^{\frac{3}{2}}} \bigg],
\label{FW-exact}
\end{equation} 
for $ |h_1| + |h_2| < a$.

\subsubsection{Discussion.}
Plots analogous to the ones showed in \fref{leading2ntol} can be drawn for the case of weak-coupling. However, since these have the same physical content as those for the Dirichlet case we do not show them here. Instead we are going to compare the perturbation expansion and the PFA to the exact weak-coupling solution. 

In order to do that, we follow the notation in \eref{weight-factor-D} and normalize the force for the weak-coupling as
\begin{equation}
{\cal F}_{W} \left( k_0a,\frac{h}{a},k_0y_0 \right)
= \frac{F_{\rm{Lat,W}} \left(k_0a,\frac{h}{a},k_0y_0 \right)}
       {\left| F^{(0)}_{\rm{Cas,W}} \right| 
        \,\frac{h^2}{a^2} \,k_0a \,\sin (k_0y_0)}.
\end{equation}
The entire expression up to fourth order in the perturbation series in the weak coupling reads
\begin{equation}
{\cal F}_{W}^{[\infty,4]}
= \frac{e_2(k_0a)}{e^{k_0a}} + 3 \frac{h^2}{a^2}
\bigg[ \frac{e_4(k_0a)}{e^{k_0a}} 
- \cos (k_0y_0) \,\frac{e_4(2k_0a)}{e^{2k_0a}} \bigg].
\end{equation}
The result in the proximity force approximation should be valid to the
first order in $k_0a$ and any order in $h/a$. Thus ${\cal F}^{[1,\infty]}$ will represent 
the PFA result in our notation. In the weak coupling this is
\begin{equation}
{\cal F}_{W}^{[1,\infty]} = \frac{1}{(1-\frac{r^2}{a^2})^\frac{3}{2}},
\end{equation}
and for the exact calculation from \eref{FW-exact} gives
\begin{equation}
{\cal F}_W^{[\infty,\infty]} =
- \frac{(k_0a)^3}{\sin (k_0y_0)} \frac{1}{\pi} 
\int_{-\infty}^\infty \frac{dt}{t}
\,{\rm Re} \bigg[ \frac{\sin (t + k_0y_0)}
            {[(t + \rmi k_0a)^2 + \{k_0 r(t)\}^2]^\frac{3}{2}} \bigg],
\quad \rm{for} \quad 2h < a.
\label{FW=intt}
\end{equation}
The above non-perturbative expression for the 
lateral Casimir force on corrugated 
plates (in the weak-coupling limit), has been plotted in 
\fref{exact-W-fig}. The perturbative 
results are plotted as dashed curves in the same figure.
We observe that the perturbative results,
when the next-to-leading order is included, 
compares with the exact result remarkably well for 
$k_0h \ll 1$ keeping the condition that $2h < a$.
The fractional error in the lateral force when the next-to-leading
order contribution is included is displayed on the right column in
\fref{exact-W-fig}. For a fixed offset, the error gets larger for higher values of $k_0h$. For $k_0h$ small, we get very good approximation in the whole range of $k_0a$.

We can also compare the exact result with the one obtained from the PFA.
\begin{figure}
\begin{tabular}{cc}
\includegraphics[width=75mm]{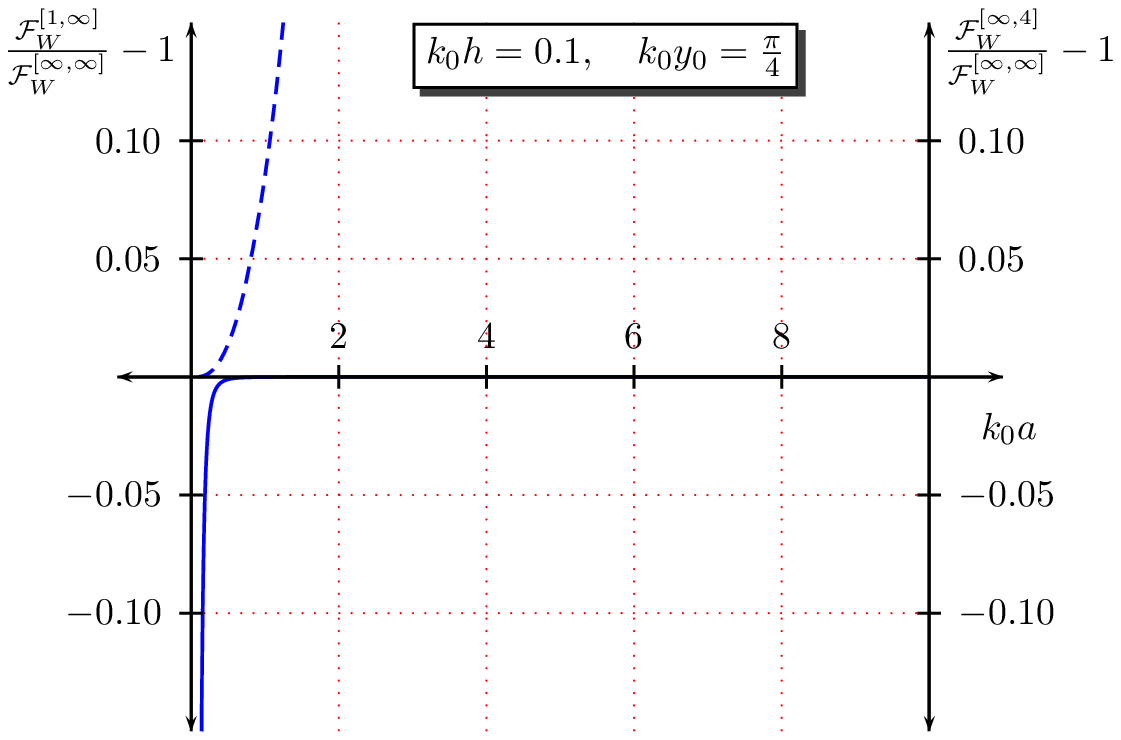} &
\hspace{5mm}
\includegraphics[width=75mm]{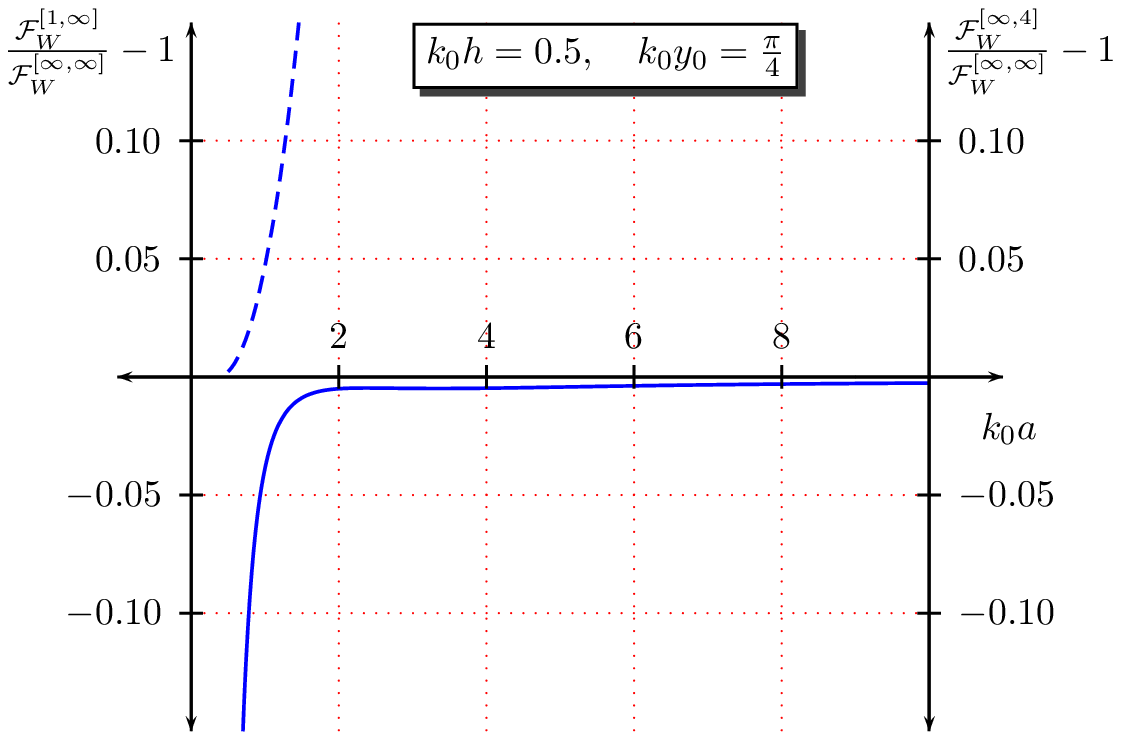} 
\end{tabular}
\caption{ Weak coupling limit:
Fractional error in PFA is plotted as the dashed curve.
The corresponding error in the perturbative result is plotted 
as the bold curve and is described by the axis on the right.}
\label{pfa-and-exact-W}
\end{figure}
In \fref{pfa-and-exact-W} we plot the fractional error 
in the PFA result, for fixed $k_0h$, and compare it to the 
fractional error in the perturbative results.
We note that PFA is a good approximation for $k_0a \ll 1$ 
and perturbative results are valid for $k_0h \ll 1$.
Both the PFA and the perturbative analysis are restricted to $2h<a$.

\begin{figure}
\begin{tabular}{cc}
\includegraphics[width=80mm]{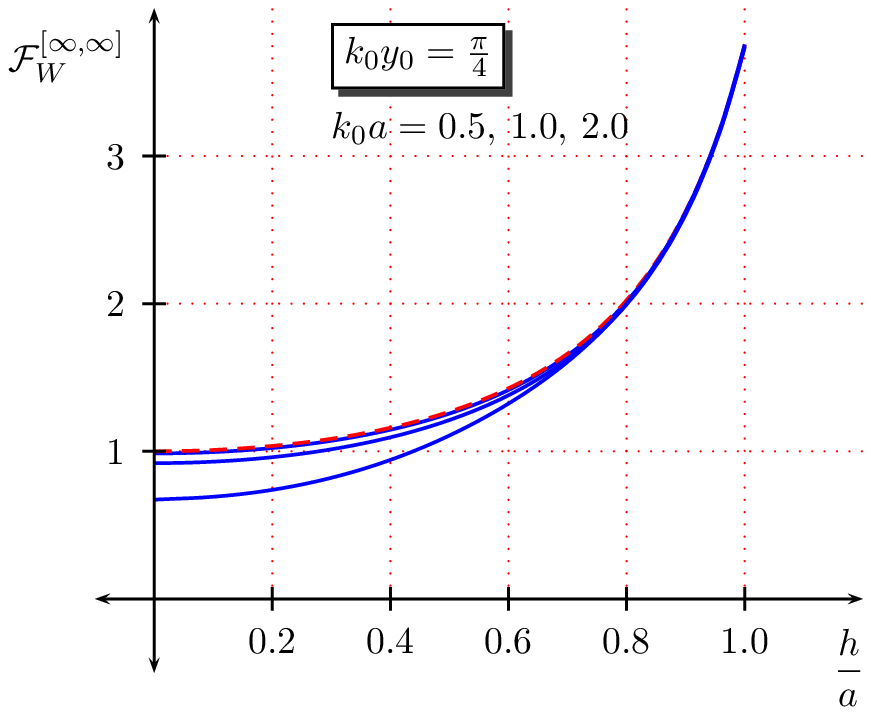} &
\hspace{5mm}
\includegraphics[width=80mm]{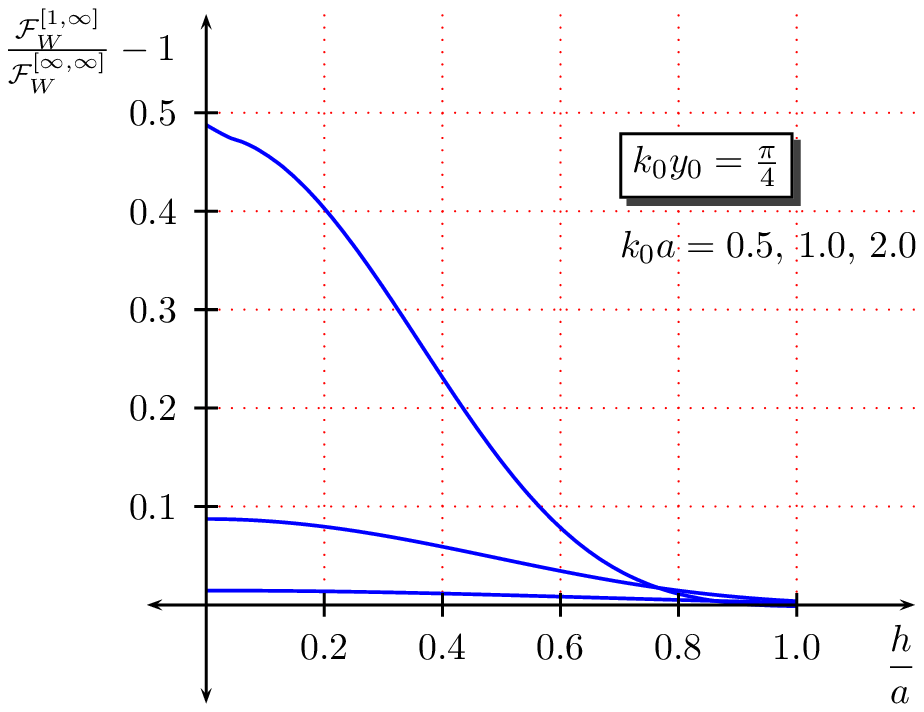}
\end{tabular}
\caption{ Weak coupling limit:
On the left the PFA is shown as the dashed curve. The exact results for 
higher values of $k_0a$ are seen to deviate from the dashed curve.
On the right the fractional error in PFA is shown to increase
with higher values of $k_0a$.}
\label{pfa-and-exact-W-versus-hoa}
\end{figure}

We earlier noted that the PFA limit is obtained by taking the
limit $k_0a \rightarrow 0$ while keeping the ratio 
$h/a$ fixed. In \fref{pfa-and-exact-W-versus-hoa}
we plot the lateral force in the PFA limit and compare it 
with the exact result for various values of $k_0a$.
We note that the error in the PFA is less than 1\% for $k_0a \ll 1$
for arbitrary $h/a$.
For $k_0y_0 = \pi/4$, we observe that the PFA is a very good
approximation for $h\sim a$ which satisfies $2h\sin (k_0y_0/2) < a$.
After viewing the plots for various values of $k_0y_0$ we note that 
in general the PFA is a good approximation for $h\sim a$ and 
further beyond for offsets $k_0y_0 < \pi/4$.
It is, in fact, plausible that the PFA holds for large amplitude
corrugations for small offsets because the corrugations fit together
like fingers in a glove.

The experiments are highly grounded on calculations based on the proximity force approximation. There has been some misunderstanding regarding the range of validity of the PFA versus numerical calculations~\cite{rodrigues2006,chen2007,rodrigues2007}. As we have shown here the ranges of validity of both solutions may not overlap.


\section{Concentric cylinders with corrugations.}
\begin{figure}
\begin{center}
\includegraphics[width=50mm]{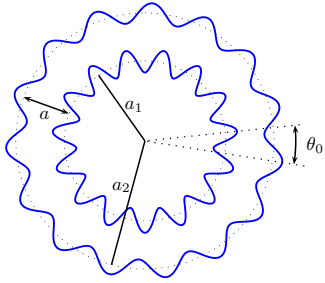}
\caption{Non-contact gears: Concentric corrugated cylinders 
with the same corrugation frequency, $\nu=15$, on each cylinder. 
$\theta_0$ is the angular shift between the gears. }
\label{corru-cyl}
\end{center}
\end{figure}
In this section we show the numerical calculation for the leading-order Casimir torque between two concentric corrugated cylinders. The cylinders have average radii $a_1$ and $a_2$ and the mean distance between them is $a=a_2-a_1$ like it is shown in \fref{corru-cyl}. The set-up is equivalent to the one in the previous case. The cylinders are described by a potential of the same kind as before,
\begin{equation}
V_i(r,\theta) = \lambda_i \,\delta (r - a_i - h_i(\theta)),
\end{equation}
where $i=1,2$ identify the individual cylinders and $h_i(\theta)$ are the functions describing the corrugations. 
For an initial angular shift $\theta_0$ as shown in \fref{corru-cyl}, and assuming sinusoidal corrugations, we have
\numparts
\begin{eqnarray}
h_1(\theta) &=& h_1 \sin [\nu(\theta + \theta_0)],
\\
h_2(\theta) &=& h_2 \sin [\nu \theta],
\end{eqnarray}
\endnumparts
where $h_{1,2}$ are the corrugation amplitudes and
$\nu$ is the frequency associated with the 
corrugations.

In the next sections we will show the results for the strong and weak coupling limit up to second order perturbation. The Casimir torque is calculated by taking the derivative of the interaction energy with respect to the shifted angle,
\begin{equation}
{\cal T} = - \frac{\partial E_{12}}{\partial \theta_0}.
\label{torque}
\end{equation}
Details of the calculation can be found in reference \cite{gearsII, AFcas}.
In the last section we discuss the results.


\subsection{Dirichlet limit.}
In the Dirichlet limit the interaction energy
can be expressed in the form
\begin{equation}
\frac{E_{12}^{(2)}}{2\pi R\,L_z}
= \cos(\nu\theta_0) \,\frac{\pi^2}{240 \,a^3} 
\,\frac{h_1}{a} \frac{h_2}{a} \, B_\nu^{(2)D}(\alpha),
\label{DE12-D-cyl}
\end{equation}
where we have divided by a factor of $2\pi R$, which is the 
mean circumference. The function $B_\nu^{(2)D}$ is defined as
\begin{equation}
B_\nu^{(2)D} (\alpha)
= \frac{15}{\pi^4} \sum_{m=-\infty}^{\infty} 8 \alpha^3 \int_0^\infty x \,dx
\frac{4 \, \alpha^2}{(1 - \alpha^2)}
\frac{1}{D_m(\alpha;x)}
\frac{1}{D_{m+\nu}(\alpha;x)},
\end{equation}
where $\alpha=\frac{a}{2R}$, $R=\frac{a_1+a_2}{2}$ and the functions $D_m$ are given by
\begin{equation}
D_m(\alpha;x) = I_m\big(x[1 + \alpha]\big) K_m\big(x[1 - \alpha]\big)
- I_m\big(x[1 - \alpha]\big) K_m\big(x[1 + \alpha]\big),
\end{equation}
\begin{figure}
\begin{center}
\includegraphics[width=80mm]{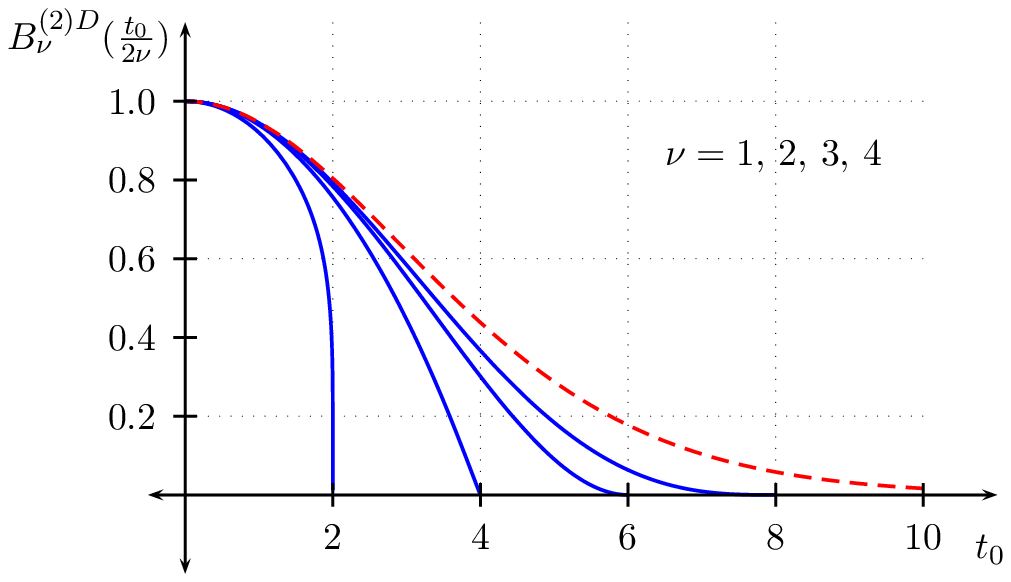}
\caption{
Dirichlet limit: Plots of $B_\nu^{(2)D}(\frac{t_0}{2\nu})$ versus $t_0$,
for $t_0 < 2\nu$ and fixed $\nu$.
The dashed curve is the corresponding plot for corrugated plates
which is approached by the corrugated cylinders for larger values
of $\nu$. }
\label{diri-fig}
\end{center}
\end{figure}
Using \eref{torque} we evaluate the Casimir torque to be
\begin{equation}
\frac{{\cal T}^{(2)D}}{2\pi R\,L_z}
= \nu \sin(\nu\theta_0) \,\frac{\pi^2}{240 \,a^3} 
\,\frac{h_1}{a} \frac{h_2}{a} \, B_\nu^{(2)D}(\alpha),
\end{equation}
where $B_\nu^{(2)D}(\alpha)$ behaves as in \fref{diri-fig}.


\subsection{Weak coupling limit.}
According to reference \cite{gearsII} the interaction energy in the limit $a\lambda_i\ll 1$ becomes,
\begin{equation}
\frac{E_{12}^{(2)W}}{2\pi R\,L_z}
= \cos(\nu\theta_0) \, \frac{\lambda_1 \lambda_2}{32 \pi^2\,a}
\,\frac{h_1}{a} \frac{h_2}{a} \, B_\nu^{(2)W}(\alpha),
\label{DE12-W-cyl}
\end{equation}
where we have defined the function
\begin{equation}
B_\nu^{(2)W}(\alpha)
= - \frac{\alpha^3}{2} \frac{\partial}{\partial \alpha}
\bigg[ \frac{1}{\alpha^2} \left( \frac{1-\alpha}{1+\alpha} \right)^\nu
(1-\alpha^2) (1 + 2\alpha\nu + \alpha^2) \bigg].
\label{bnu-W}
\end{equation}
The Casimir torque per unit area, for the weak coupling limit, can thus
be evaluated to be
\begin{equation}
\frac{{\cal T}^{(2)W}}{2\pi R\,L_z}
= \nu \sin(\nu\theta_0) \, \frac{\lambda_1 \lambda_2}{32 \pi^2\,a}
\,\frac{h_1}{a} \frac{h_2}{a} \, B_\nu^{(2)W}(\alpha).
\end{equation}
\begin{figure}
\begin{center}
\includegraphics[width=80mm]{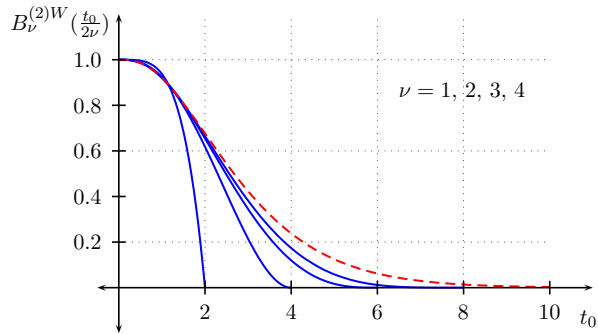}
\caption{
Weak coupling limit: Plots of $B_\nu^{(2)W}(\frac{t_0}{2\nu})$ versus $t_0$,
for $t_0 < 2\nu$ and fixed $\nu$.
The dashed curve is the corresponding plot for corrugated plates
which is approached by the corrugated cylinders for larger values
of $\nu$. }
\label{weak-fig}
\end{center}
\end{figure}
The function $B_\nu^{(2)W}(\alpha)$ is plotted in \fref{weak-fig}.


\subsection{Discussion.}

We can compare our results with the results from the corrugated parallel plates. In the limiting case when the two cylinders have very large radii but the distance between them is kept constant we can consider that we have parallel plates  in a region of small variations in the angle $\theta_0$. This corresponds to $m,n\rightarrow\infty$ such that $m/R$ is finite and $\alpha\rightarrow 0$.

In \fref{diri-fig} we plot the function $B_\nu^{(2)D}(\alpha)$, where $t_0=2\alpha\nu$. The dashed curve corresponds to the parallel plates. We see that the plot approaches the result for parallel plates as we increase $\nu$. Notice that $\alpha$ cannot get values bigger than $1$ since that case corresponds to the inner radius equal to $0$. When $\alpha\rightarrow 0$ we reach the PFA limit.

The same conclusions can be drawn from \fref{weak-fig} for the weak coupling case.

In general we see that our results for the Casimir torque reproduce the results for the lateral force on corrugated parallel plates in the limit of large radius and small corrugation wavelengths.


\ack
We thank the US National Science Foundation, grant number PHY-0554826,
and the US Department of Energy, grant number DE-FG-04ER41305, for
partial support of the research reported here. ICP would like to thank the French National Research Agency (ANR)
for support through Carnot funding. We show our gratitude to
Victor Dodonov for his hospitality and his great organization of this meeting on ``60 Years
of the Casimir Effect.'' We also like to thank the International Center for Condensed
Matter Physics at the University of Brasilia where the meeting was held.


\end{document}